\documentclass[aps,twocolumn,pre,noshowpacs,superscriptaddress,noshowkeys,floatfix]{revtex4}

\usepackage{graphicx}
\usepackage{amsthm,amsmath}
\usepackage{natbib}

\begin{document}

\title{Extrusion without a motor: a new take on the loop extrusion model of genome organization}

\author{C. A. Brackley}
\author{J. Johnson}
\author{D. Michieletto}
\author{A. N. Morozov}
\affiliation{SUPA, School of Physics and Astronomy, University of Edinburgh, Peter Guthrie Tait Road, Edinburgh, EH9 3FD, UK}
\author{M. Nicodemi}
\affiliation{Dipartimento di Fisica, Universita' di Napoli Federico II, INFN Napoli, CNR, SPIN, Complesso Universitario di Monte Sant'Angelo, 80126 Naples, Italy}
\author{P. R. Cook}
\affiliation{Sir William Dunn School of Pathology, University of Oxford, South Parks Road, Oxford, OX1 3RE, UK}
\author{D. Marenduzzo}
\affiliation{SUPA, School of Physics and Astronomy, University of Edinburgh, Peter Guthrie Tait Road, Edinburgh, EH9 3FD, UK}

\begin{abstract}
Chromatin loop extrusion is a popular model for the formation of CTCF loops and topological domains. Recent HiC data have revealed a strong bias in favour of a particular arrangement of the CTCF binding motifs that stabilize loops, and extrusion is the only model to date which can explain this. However, the model requires a motor to generate the loops, and although cohesin is a strong candidate for the extruding factor, a suitable motor protein (or a motor activity in cohesin itself) has yet to be found. Here we explore a new hypothesis: that there is no motor, and thermal motion within the nucleus drives extrusion. Using theoretical modelling and computer simulations we ask whether such diffusive extrusion could feasibly generate loops. Our simulations uncover an interesting ratchet effect (where an osmotic pressure promotes loop growth), and suggest, by comparison to recent \textit{in vitro} and \textit{in vivo} measurements, that diffusive extrusion can in principle generate loops of the size observed in the data.
\\[1em]
Extra View on : C. A. Brackley, J. Johnson, D. Michieletto, A. N. Morozov, M. Nicodemi, P. R. Cook, and D. Marenduzzo ``Non-equilibrium chromosome looping via molecular slip-links'', Physical Review Letters \textbf{119} 138101 (2017)\\[1em] 
\end{abstract}

\maketitle

The development of a high-throughput version of chromosome conformation capture experiments (HiC) has led to some paradigm-shifting discoveries about the three-dimensional (3D) organisation of chromosomes within the nucleus. First, it was found that the genome can be split into two ``compartments''~\cite{LiebermanAiden2009}, where active chromatin preferentially interacts with other active regions, and inactive chromatin preferentially interacts with other inactive regions. Active and inactive regions are normally called A and B compartments respectively. Next came the identification of topologically-associating domains, or ``TADs''~\cite{Dixon2012,Sexton2012}. A TAD is defined in a HiC contact map as a genomic block in which interactions between loci within a block are enriched compared to those between loci in neighbouring blocks. More recently~\cite{Rao2014}, HiC experiments have led to the discovery of ``loop domains'', which are a subset of TADs that are enclosed within a loop (i.e., there is a direct interaction between the two boundaries). Such loops are normally anchored by the  CCCTC-binding factor (CTCF) bound at its cognate sites.

The CTCF loops have an intriguing property. As the CTCF binding motif is non-palindromic, it has a direction on the DNA and can be thought of as an arrow pointing along the chromatin fibre. A pair of sites on the same chromosome can therefore be in one of four possible arrangements: the two motifs could point towards each other or away from each other, both could point forward, or both could point backwards. High-resolution HiC data~\cite{Rao2014} revealed that in over 90\% of CTCF loops, the motifs point towards each other (they are convergent). This striking observation is difficult to explain, as it requires that large scale information on the nature of a genomic loop is somehow transmitted to a protein complex containing CTCF. A simple picture of loop formation might entail a thermodynamic model where two CTCF sites come into contact through random 3D diffusion, and then stick together thanks to some biochemical affinity. But then how could such a pair of sites ``know'' about the large scale arrangement, and determine whether they should bind or not?

One way in which information about genome organisation could be transferred along the chromosome, is through a tracking mechanism where a protein binds at one point, and then tracks along the chromatin to reach another. The loop extrusion model is a popular idea proposed by several groups~\cite{Nasmyth2001,Alipour2012,Sanborn2015,Fudenberg2016} to explain the CTCF bias: some loop-extruding factor binds to the chromatin at a single point, folds it into a loop, and then tracks along it in opposite directions to grow, or ``extrude'', this loop. Thus, information about the direction of the loop is transmitted down the fibre; in the model the factor is halted when it meets a CTCF bound to a site with its motif oriented towards it, but continues extruding if it meets a CTCF pointing the other way. This naturally explains the looping bias.

Loop extrusion is an appealing model: as well as explaining the motif orientation bias, computer simulations have shown that it can also give a very good prediction of the TAD structure observed in HiC data (though this requires a constant flux of extruders and depends on the choice of parameters~\cite{Fudenberg2016}). Since the motif bias was first discovered, disruption of CTCF binding (using genome editing to remove, or even reverse the orientation of binding sites) has been shown to alter domain organisation and affect promoter-enhancer interactions~\cite{DeWit2015,Sanborn2015,Guo2015}: loop extrusion can successfully predict many of these observations. A strong candidate for the loop-extruding factor is the SMC complex cohesin~\cite{Uhlmann2016}. This ring-like complex topologically embraces DNA and chromatin~\cite{Ivanov2005,Stigler2016,Davidson2016}, and is found with CTCF at loop anchors (showing a bias to be found to one side of the CTCF motif -- justifying the assumption that the interaction between CTCF and extruders is directional). Additionally, cohesin has been observed to translocate away from its loading sites to become enriched elsewhere~\cite{Lengronne2004,Busslinger2017}. 

While extrusion can seemingly predict many of the interaction patterns observed in HiC data, the idea remains controversial. One crucial requirement of the model is a motor activity, needed to push cohesin and generate the loop. Which protein is the motor? How much biochemical energy is required? How is the direction of extrusion maintained to promote loop growth (and not shrinking)? These are all as yet unanswered questions. Though cohesin does have an ATPase activity, this is thought to be involved in ring opening and closing, and not directional motion. Interestingly, the related condensin complex has recently been shown to be able move unidirectionally along DNA in the presence of ATP~\cite{Terakawa2017}, but under similar conditions cohesin only shows diffusive motion~\cite{Davidson2016,Kanke2016}. An alternative possibility is that cohesin is pushed by another motor. In any case the motor must generate loops of 100-1000~kbp within the residence time of cohesin on chromatin (about 20-25~min~\cite{Gerlich2006,Ladurner2014,Hansen2017}). This means that the motor must travel, at the very least, at speeds of 2-20~kbp/min. While some bacterial translocases can travel even faster, the required speed far outstrips that of RNA polymerase (1~kbp/min) which is one of the most processive motors active in interphase.

The loop extrusion hypothesis has inspired many recent publications on CTCF and cohesin, so it would seem that the search is on for the mystery motor which does the extrusion. Here, however, we consider an alternative. What if there is {\it no motor at all}? What if a cohesin ring encircles the chromatin fibre in a way that allows it to diffuse freely along that fibre? Can the thermal energy in the nucleus provide enough diffusive motion to extrude loops without a motor? We explore this possibility using theoretical modelling, computer simulations, and the latest \textit{in vitro} and \textit{in vivo} data.

\section*{Diffusive extrusion: a non-equilibrium model.}

\begin{figure}
\includegraphics[width=0.5\textwidth]{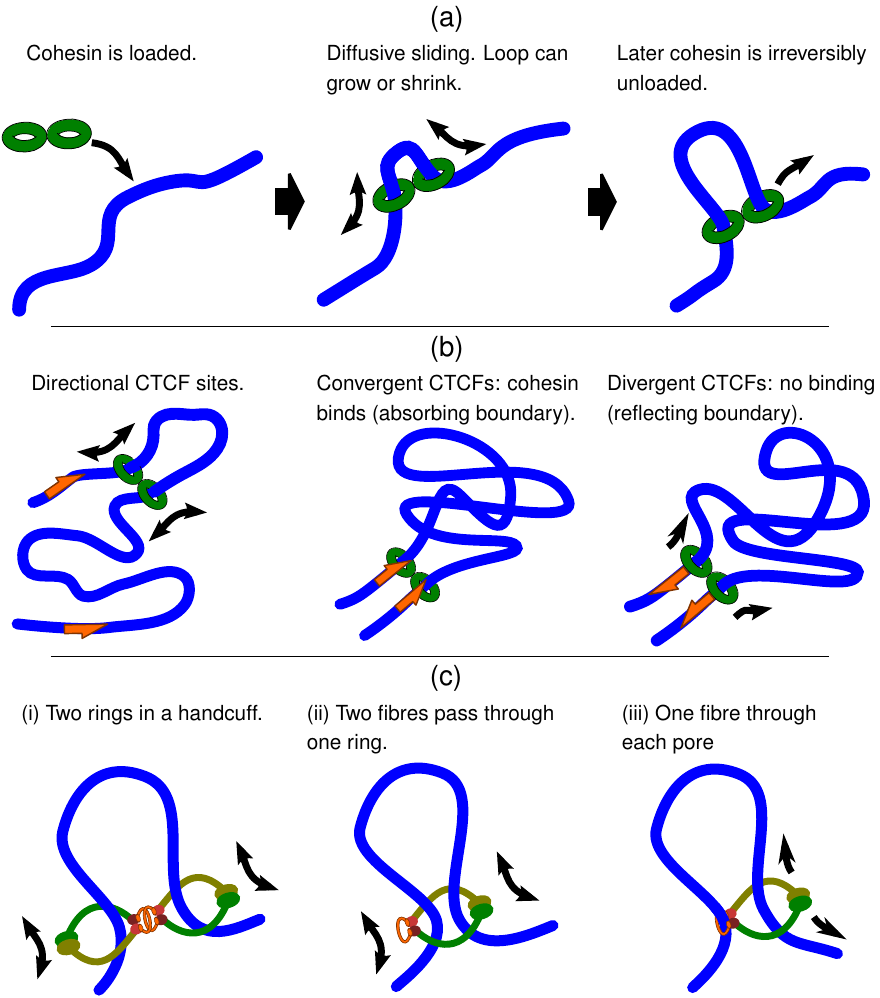}
\caption{Cartoon describing the diffusive loop extrusion model. (a) Cohesin is loaded onto chromatin fibre at two adjacent points. Here a pair of cohesins is shown as a handcuff. The cohesin and chromatin are then able to diffuse such that the rings slide along the fibre; a loop can grow and shrink. Later the cohesin is unloaded; since loading can only occur at adjacent positions, but unloading can occur while there is a loop, the process is not time reversible. The handcuff is unloaded stochastically with rate $k_{\rm off}$, and an unbound handcuff is reloaded with rate $k_{\rm on}$. This geometry is the same as the active loop extrusion case, but no motor action is required to grow the loop. (b) If cohesin interacts directionally with CTCF, binding only when it is pointing towards it, then convergent CTCFs form an absorbing boundary whereas divergent CTCFs form a reflecting boundary. Only for the convergent orientation will a stable CTCF loop form, in agreement with HiC experiments. (c) Cartoons showing alternative cohesin loading configurations which could accommodate diffusive loop extrusion: (i) shows a pair of cohesins as a handcuff; (ii) and (iii) show possible configurations for a single cohesin ring.
\label{fig:cartoons}}
\end{figure}

We consider a simple picture where a pair of cohesin complexes are loaded at adjacent positions on a chromatin fibre in a handcuff configuration [Fig.~\ref{fig:cartoons}(a)]. This is one easy-to-visualize arrangement -- everything below also holds for a single ring encircling the fibre at two points [various alternative arrangements are shown in Fig~\ref{fig:cartoons}(c)]. We then assume that the handcuff can diffuse by sliding along the fibre(s), and a loop will grow and shrink diffusively. Then, sometime later, the cohesin will be unloaded from the chromatin. Importantly, even though the motion is diffusive, this is still a ``non-equilibrium'', or active system. In the language of statistical physics, detailed balance is broken since cohesin is always loaded at adjacent points on the chromatin, a loop can grow or shrink, and cohesin can be unloaded (but not loaded) where there is a finite-sized loop (i.e., the system is not time reversible). Biologically, it is thought that chemical energy is required both to load and unload cohesin from the fibre (requiring both ATP hydrolysis and specific loading/unloading factors~\cite{Murayama2015}): this provides a mechanistic justification for considering a non-equilibrium model. If, when a diffusing cohesin meets a DNA-bound CTCF protein, it either forms a complex with CTCF or it reflects off it (i.e. just diffuses away again) depending on the CTCF orientation, then this explains the bias for convergent CTCF motifs in loops. Diffusive extrusion is in many ways similar to the active extrusion model discussed above. 

Now the question is whether diffusion can generate loops of the required size within the allowed time (the mean residence time of cohesin on DNA). In the active extrusion case the motor would either have to track along the DNA contour (negotiating nucleosomes and other obstacles along the way), or it would have to step along the nucleosomal fibre while maintaining a fixed direction of motion. In the diffusive case, the cohesin ring instead diffuses over whatever fibre structure is present \textit{in vivo}. The important quantities are therefore the effective diffusion constant for 1D motion along the fibre, and the linear compaction of that fibre [e.g., the number of bp per nanometre (nm)]. A simple theoretical model (full details are given in Ref.~\cite{PRL}) can put some limits on what these quantities can be in order that diffusive extrusion is viable. For example if we need to generate 100~kbp loops within 25~min, the theory tells us that a 1D diffusion constant of at least 10~kbp$^2$/s is required: if chromatin exists as a 30~nm fibre with about 100~bp/nm, this equates to $D\sim$0.001$\mu$m$^2$/s as a minimum diffusion constant. If a more conservative estimate of 20~bp/nm is used (corresponding to a relatively open fibre), then diffusive extrusion is viable if $D\sim$0.025$\mu$m$^2$/s or above. Recent \textit{in vitro} experiments of acetylated cohesin diffusing on chromatin fibres reconstituted in \textit{Xenopus} egg extract found $D=0.2525 \pm 0.0031 \mu$m$^2$/s; although this was on stretched chromatin in a dilute solution, if the \textit{in vivo} value is anywhere near this, diffusive extrusion may well be feasible. Other recent \textit{in vitro} work~\cite{Stigler2016} studied cohesin on DNA with nucleosome-like obstacles: they found that cohesin did not translocate over obstacles larger than 20~nm, and extrapolating crossing times for smaller obstacles suggested that cohesin would be able to travel 7~kbp in 1 hour (this would correspond to $D=0.0003$ $\mu$m$^2$/s and a compaction of 3.4~bp/nm, only suitable for naked DNA, hence this extrapolation is in practice a lower bound). If diffusion \textit{in vivo} is closer to that estimate, then diffusive extrusion would seem less feasible (but see below).

\begin{figure*}
\centering
\includegraphics{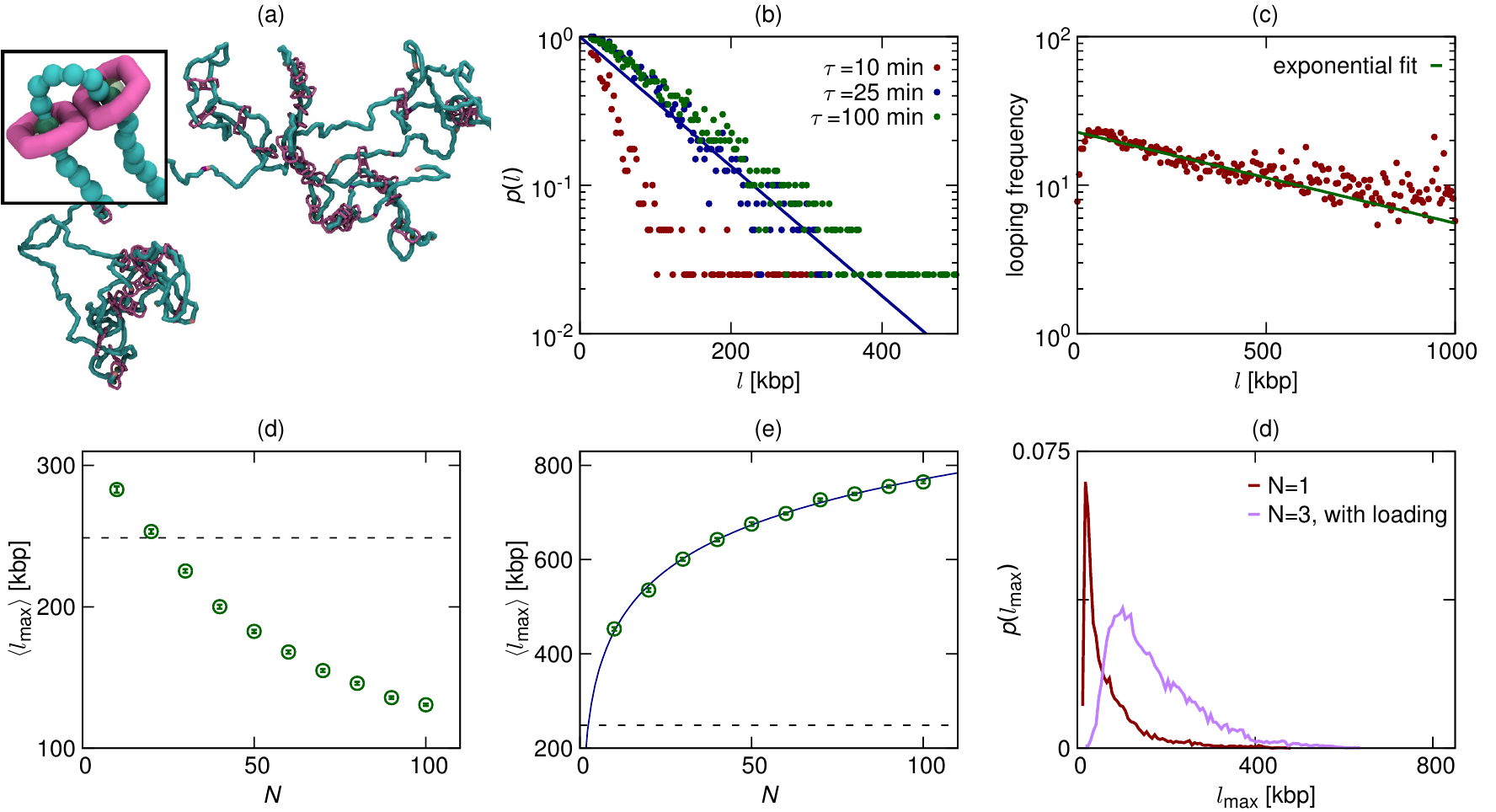}
\caption{Simulations and theory for diffusive loop extrusion. (a) Snapshot of a 3D Brownian dynamics simulation in which multiple handcuffs bind/unbind from a 4.5~Mbp fibre. The inset shows a zoom of one handcuff. 
(b) The probability that diffusive loop extrusion generates a loop of size $l$ is obtained from simulations such as that shown in (a) but with a single handcuff, for different values of the unloading rate $k_{\rm off}=\tau^{-1}$. The solid line shows an exponential fit for $\tau=25$~min.
(c) Plot showing the frequency at which loops of different sizes are observed in ChIA-PET CTCF pull-down (data from Ref.~\cite{Tang2015}). This fits better to an exponential function (green line) than a power law.
(d-e) Plots showing how the mean size of the largest loop in a simplified 1D simulation depends on the number of cohesin handcuffs for two loading scenarios (see Ref.~\cite{PRL} for details). In (d) handcuffs are loaded at a randomly chosen site on the chromatin each time they bind, whereas in (e) handcuffs are always loaded at the same site. Dashed lines indicate the loop size for the case of a single handcuff; the solid line in (e) shows a fit to the equation $a+b\log(N)$ which is the functional dependence on $N$ predicted by the theory.
(f) Plot showing the distribution of the size of the largest loop in 3D simulations for $N=1$ and $N=3$ cohesin handcuffs, for the case where loading is always at the same site. The stark difference illustrates that the ratchet effect is in operation even for a small number of nested handcuffs.
\label{fig:ratchet} }
\end{figure*}

\section*{3D simulations of diffusing extruders}

As well as theoretical modelling, we also performed 3D Brownian dynamics simulations (full details are given in Ref.~\cite{PRL}) to assess whether diffusive extrusion can generate loops, rearranging large stretches of chromatin within the crowded nuclear environment. In these polymer-physics based simulations (which are similar to those in previous studies~\cite{Brackley2013,Barbieri2012,Brackley2016}, but with some additions described below) the chromatin fibre is represented as a simple chain of beads connected by springs. Each bead represents 3~kbp of chromatin (though similar results are obtained with different values), and we simulate stochastic diffusive motion of the chain. In previous works on the active extrusion model~\cite{Sanborn2015,Fudenberg2016}, extruding factors were represented by extra springs which move actively along the fibre. Here we explicitly simulate a pair of molecular handcuffs (made up of beads similar to the chromatin) which can slide diffusively on the chromatin. The handcuffs are attached to, and removed from, the fibre at time intervals according to a Poisson process (having a mean residence time $\tau=k_{\rm off}^{-1}$); they are always loaded as a pair onto two adjacent chromatin beads. These dynamics mimic an active, ATP-dependant, loading-unloading process which drives the system away from equilibrium. Figure~\ref{fig:ratchet}(a) shows a snapshot of part of a simulated fibre.

Figure~\ref{fig:ratchet}(b) shows a plot of the probability that the simulation generates a loop of a given size, for different values of the unloading rate. There is a significant probability of finding loops of several hundred kbp, implying that the diffusive extrusion mechanism is likely capable of the rearrangement of the chromatin fibre necessary to form a loop. The diffusion constant for cohesin sliding is $D=2.3\times10^{-3}~\mu$m$^2$s$^{-1}$ (which arises naturally from the geometry of our simple bead-based model); this is much smaller than the \textit{in vitro} value for chromatin in \textit{Xenopus} extract quoted above, so the simulations provide a conservative estimate of the probability to form loops (still, this value is sufficient to create large loops diffusively). 

An interesting feature of the plot in Fig.~\ref{fig:ratchet}(b) is that the probability of forming a loop is approximately an exponential function of loop length (theoretical modelling predicts an exponential decay with a power-law correction). Standard equilibrium polymer physics would predict that the probability of forming a loop of length $l$ is a simple power-law function of $l$~\cite{deGennes}, so the non-equilibrium binding/un-binding kinetics have indeed altered the looping behaviour. HiC data shows that \textit{in vivo} the probability of two loci interacting decreases with a power-law function of their genomic separation on average~\cite{LiebermanAiden2009}; however, ChIA-PET data~\cite{Tang2015} obtained using an antibody targeting CTCF (which therefore only includes interactions between CTCF bound loci), fit better to an exponential decay [see Fig.~\ref{fig:ratchet}(c)]. Though there are likely many other factors affecting these data, this suggests that different mechanisms are at play for CTCF loop formation to those behind chromosome interactions in general. 

\section*{A ratchet effect promotes loop growth over shrinking}

In the simulations and theory discussed so far we have considered only a single bound cohesin handcuff, whereas \textit{in vivo} we might expect many bound cohesins to from a complicated pattern of loops. In the active loop extrusion simulations presented in Refs.~\cite{Fudenberg2016,Sanborn2015} there are many extruding factors which bind at random locations throughout the genome. However, \textit{in vivo} the cohesin-loading factor (NIPBL in humans, or Scc2 in yeast) binds at preferred genomic locations, and there is some evidence that cohesin is loaded near the promoters of active genes~\cite{Kagey2010}. Our simulations allow us to investigate both loading at random locations and at preferred sites.

Interestingly the dynamics are very different in the two cases. Figures~\ref{fig:ratchet}(d) and (e) show results from simple 1D simulations (full details are given in Ref.~\cite{PRL}) where different numbers of handcuffs are continually being loaded and unloaded from a fibre with a mean residence time of 20~min. If the handcuffs are loaded at randomly chosen locations each time, a series of loops form side by side, competing with each other for space. The average loop size decreases as the number of loops increases. If handcuffs are loaded only at a single location, then the loops tend to be nested inside each other; this leads to an interesting ratchet effect, where the inner loops promote growth over shrinking of the outer loops. This has a simple explanation: when the first handcuff binds it follows a 1D random walk; when the second binds at the same site, it prevents the first from diffusing back towards the loading site, i.e., it exerts an osmotic pressure on the outer handcuff.  This osmotic pressure means that the size of the largest loop \textit{increases} with the number of cohesins. This ratchet effect gives a possible mechanism through which diffusive extrusion might be accelerated, meaning it could be feasible even for smaller 1D diffusion coefficients. Further 3D simulations show that the effect is at work even for a small number of handcuffs [Fig.~\ref{fig:ratchet}(f)]. Another recent work~\cite{Yamamoto2017} proposed that the osmotic pressure can be further enhanced by interspersing pairs of cohesin complexes arranged as handcuffs with single cohesins which diffuse along the fibre but are not linked at multiple points so do not form loops.

\section*{Domains with diffusive extruders}

\begin{figure}
\includegraphics{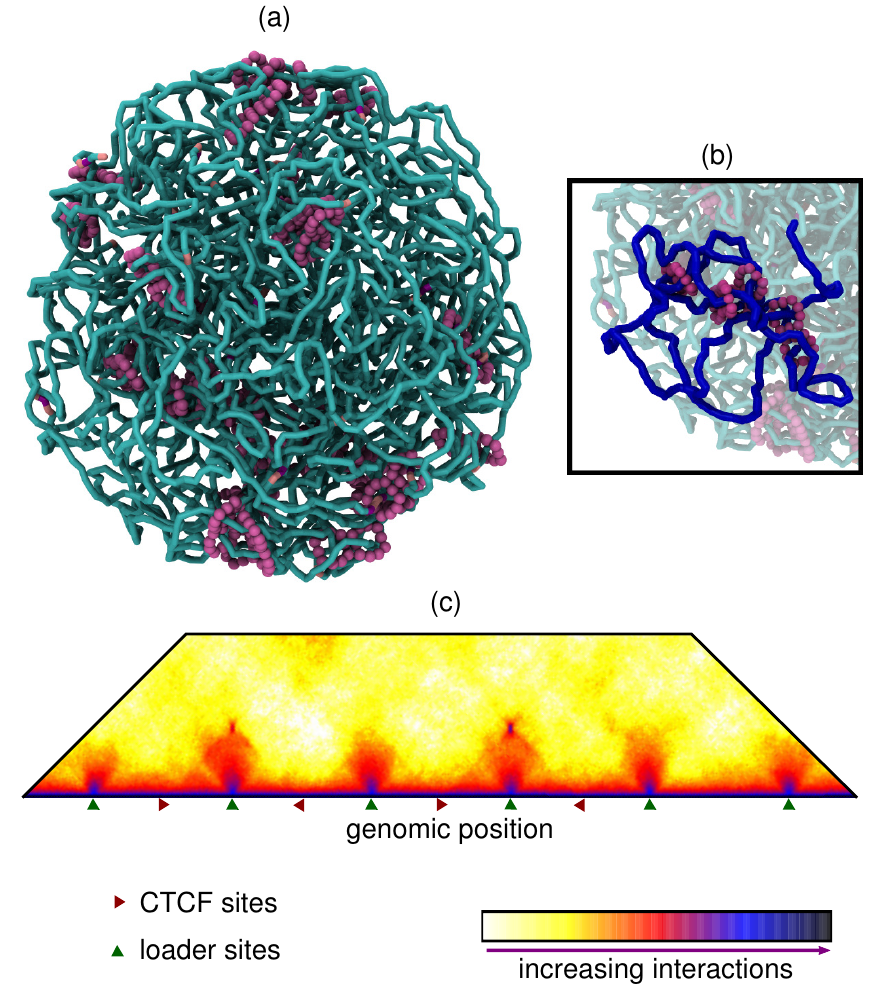}
\caption{Large scale 3D simulation. (a) Simulation snapshot of large polymer representing a 15~Mbp chromatin fibre, with 32 diffusing handcuffs. Here the polymer is confined within a sphere so as to give a realistic chromatin density (using periodic boundaries with the same density instead of confinement gives similar results). (b) Zoom of the same snapshot, but with one domain highlighted in blue. 
(c) A HiC-like interaction map is shown for a 300~kbp region of the simulated fibre. The colour at each point in the map indicates the frequency of interaction between the chromatin positions connected by a triangle with its apex at that point. Positions and orientations of CTCF sites, and positions of the loader sites are indicated.
\label{fig:domains} }
\end{figure}

Large-scale Brownian dynamics simulations can also be used to investigate whether diffusing cohesins can generate domains and interaction patterns similar to those seen in HiC data. We performed a simulation of a 15~Mbp region with realistic chromatin density; 32 pairs of handcuffs were continuously added and removed from the fibre with 16 preferred loading sites, and a mean residence time equivalent to 25~min. CTCF sites were placed at 750 or 1500~kbp intervals in either convergent or divergent arrangements. Eight repeat simulations were performed, with CTCF sites populated stochastically such that each simulation could have a different set of sites (to model cell-to-cell variation in CTCF binding); diffusing handcuffs only stick at CTCFs pointing towards them, and when there is a CTCF bound at each side of the handcuff, the unbinding rate is reduced by 10 fold. Figure~\ref{fig:domains}(a) shows a snapshot of the simulation system, and Fig.~\ref{fig:domains}(b) shows a zoom with the region between one pair of convergent CTCF sites (i.e. a TAD) highlighted in blue.

Figure~\ref{fig:domains}(c) shows a contact map generated from these simulations. As in HiC interaction maps, red triangles show domains, and dark spots are seen at the edges of convergent CTCF loop domains. Dark spots are also seen close to the diagonal at loading sites -- a feature not normally seen in HiC (though note that there remain few publicly available data sets showing genomic locations where the loader is enriched, and it has yet to be confirmed that cohesin is preferentially loaded at these sites). 

While these simulations show that some aspects of the domain structure can indeed be reproduced by our diffusion-based model, we urge caution in expecting such a simple model to be able to replicate interaction maps exactly. For example, the model does not include other DNA-binding proteins that might affect cohesin motion, nor do we attempt to account for active processes such as transcription. Elongating polymerases generate forces and torques (leading to supercoiling~\cite{Gilbert2014,Benedetti2017}) which may affect cohesin diffusion; indeed recent experiments where the WAPL cohesin unloader protein is knocked down in mouse nuclei show that cohesin collects preferentially between convergent genes, indicating that polymerase can push cohesin along the fibre~\cite{Haarhuis2017,Busslinger2017}. These caveats apply equally to the active loop extrusion model.

\section*{Discussion}

In this work we have argued that 1D diffusion of cohesin along chromatin can lead to loop extrusion {\it without} the need to invoke an explicit motor action. Of course experimental verification of this remains a significant challenge. Nevertheless, we suggest that diffusive extrusion cannot be dismissed in favour of an active extrusion model in the absence of additional experimental evidence.

The \textit{in vitro} experiments mentioned above~\cite{Stigler2016,Davidson2016} studied the topological loading and diffusion of cohesin rings on stretched DNA templates, and over obstacles. No directed motion was observed, but the diffusivity was found to strongly depend on ATPase activity, salt concentration, and on the way in which the cohesin complex was loaded onto the substrate. Diffusion on reconstituted chromatin in \textit{Xenopus} egg extracts was also measured~\cite{Kanke2016}, and it was found that acetylation of the Smc3 sub-unit strongly increased the diffusion coefficient. Together these results suggest that the pore size and diffusivity of cohesin might be regulated by ATP hydrolysis and acetylation~\textit{in vivo}. Recent \textit{in vivo} studies have shown that knocking-down the loader NIPBL (which leads to loss of chromosome-bound cohesin) leads to loss of looped domains~\cite{Schwarzer2016}, whereas a knock-down of CTCF affects intra-domain interactions~\cite{Nora2017}. All these observations are consistent with both the active and diffusive extrusion models. A third possibility is that there is some active translocation, but that the direction is not fixed and the cohesin is ``kicked'' randomly back and forth along the fibre (the overall effect would look like diffusive motion, but with an increased diffusion constant). 

Unlike cohesin, the condensin complex \textit{can} perform unidirectional active stepping along a stretched DNA template in the presence of ATP~\cite{Terakawa2017}. This points to the possibility that active extrusion may be at work during mitosis, where condensin plays a central role~\cite{Nasmyth2001,Goloborodko2016}.

Active loop extrusion is often cited as a model for the formation of topological domains, but this is not the only possible mechanism. Another popular model is that chromatin interactions are mediated by transcription factors (or complexes thereof) which can diffuse freely in 3D through the nucleus, and which are {\it multivalent}, meaning they can from molecular bridges between different genomic loci~\cite{Brackley2016,Brackley2016b,Barbieri2017}. This idea has been extensively studied using molecular dynamics and Monte Carlo simulations of simple bead-and-spring polymer physics models (sometimes referred to as the strings-and-binders-switch (SBS) model~\cite{Barbieri2012}). Using only limited data about where proteins bind (or using histone modification data to infer protein binding) it is possible to reproduce the TAD patterns observed in HiC data. For example a model using only two factors, one binding to active and one to inactive regions, correctly predicted the locations of 85\% of TAD boundaries on chromosome 19 in HUVECs (human umbilical vein endothelial cells)~\cite{Brackley2016}. This model naturally describes promoter-enhancer interactions mediated by polymerase-transcription factor complexes, or heterochromatin and polycomb repressed regions organised by HP1 and PRC complexes respectively. It can explain the formation of the domains which do not have looping between their boundaries, as well as the larger scale A/B compartment formation, and the fact that compartments are preserved upon loss of chromosome-bound cohesin or CTCF (which is difficult to reconcile with a loop extrusion model). The transcription factor model cannot, however, explain the CTCF motif bias.
 
It seems likely then, that a complete explanation of genome organisation will require a combination of loop extrusion and multivalent transcription factor models. Even so, as noted above, there are many additional processes which are not yet included in either of these models, so one should not expect to be able to reproduce, for example, all the features of a HiC interaction map. The aim of modelling and simulations therefore should not be to reproduce carbon-copies of experimental results, but should rather be to provide insight, propose new hypothesis, and help direct new experiments.

\textbf{\textit{Acknowledgements}} This work was supported by ERC (CoG 648050,THREEDCELLPHYSICS), the NIH ID 1U54DK107977-01, CINECA ISCRA Grants No. HP10CYFPS5 and HP10CRTY8P, and by the Einstein BIH Fellowship Award to MN. 


\end{document}